\documentclass[twocolumn,showpacs,preprintnumbers,amsmath,amssymb,prl]{revtex4}
\usepackage{graphicx}
\usepackage{dcolumn}
\usepackage{color}
\usepackage{bm}
\begin{document}
\title {Equilibrium study of Protein/DNA Complexes from Crystal Structure Data}
\author{ Sudipta Samanta$^{1}$, J. Chakrabarti$^{1}$ and Dhananjay Bhattacharyya$^{2}$}
\affiliation{$^{1}$S.N. Bose National Centre for Basic
Sciences,JD Block, Sector III, Salt Lake, Kolkata 700098, India.\\
$^{2}$ Biophysics Division, Saha Institute of Nuclear Physics, 1/AF Bidhannagar, Kolkata 700064, India}
\date{\today}
\begin{abstract}
From the crystal structure data and using the concept of equilbrium statistical mechanics we show how to calculate 
the thermodynamics of protein/DNA complexes. We apply the method to the TATA-box binding protein (TBP)/TATA sequence complex.
We have estimated the change in free energy and entropy for each of the base pair (bp).
The local rigidity of the DNA is estimated from the curvature of the free energy.
We also estimate the free energy gain of the protein due to bond formation with a particular bp. 
We thus figure out the bps responsible for specific binding. 
\end{abstract}
\maketitle

\begin{figure}[h]
{{\resizebox{8cm}{4cm}{\includegraphics{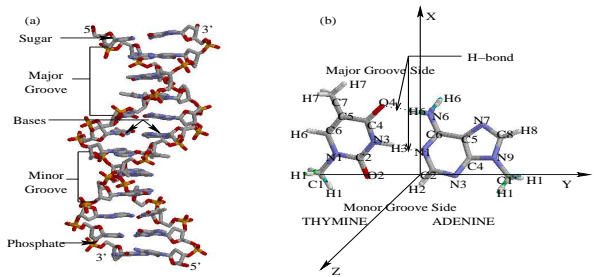}}}}
\caption{(color online) 
(a)  A nucleotide with Watson-Crick base pairing, where Adenine paired with Thymine. X and Y axes are in the plane of the base pair and Z axis is perpendicular to the base pair plane. Y axis is vector joining ${C1}^{\prime}$ atoms. 
(b) The structure of a normal DNA.}
\label{fig1}
\end{figure}

All the cellular processes in life are controlled by two biopolymers, namely, protein and nucleic acid along with many 
smaller molecules like lipid, carbohydrate, water etc. via their specific and non-specific interactions.  
The aminoacids form the primary structure of a protein molecule\cite{cell}. 
On the other hand, the strutural unit of an antiparallel double helical deoxiribo nucleic acid (DNA) molecule\cite{cell,caladine}, 
as shown in Fig.\ \ref{fig1}(a), is the 
nucleotides consisting of 5-carbon neutral sugar (deoxyribose), nitrogen-containing purine (adenine, A and guanine, G) or 
pyriminine (thymine, T and cytosine, C) hydrophobic bases attached to the sugar, the former in turn attached to a negatively charged 
phosphate group. 
The Watson-Crick base pairs (bp), namely A with T and G with C, remain at the core of the double helix with strong inter-basepair stacking, while 
the phosphates line up the periphery. Functional groups of the bases (amino, imino or keto) 
capable to form hydrogen bonds (H-bond) with functionally important protein molecules, are exposed towards the solvent within the two grooves: major or minor
as shown in Fig.\ \ref{fig1}(b).

The protein synthesis in an organism is controlled via gene expression\cite{cell}.
It consists of two parts: (i) transcription, the synthesis of ribo nucleic acid (RNA) from DNA and (ii) 
translation, the synthesis of protein from RNA. For initiation of transcription, a transcription factor
comprising a class of protein including RNA polymerase needs to bind to the specific site of DNA (promoter) that 
encodes for the given protein (gene).
Obviously, one major thrust in biochemical research aims at understanding protein/DNA interaction. 
Some of the issues of immense biochemical importance are: (i)Energetics involved in 
the complex formation; (ii) time scale related to different degrees of freedom of DNA bp in the complex formation;
(iii) kinetics of binding. 
Apart from pedagogical interest, these issues are extremly important application in drug designing, macromolecule recognition etc.

In this paper we show how the x-ray crystal structure data can be utilized to 
understand the thermodynamics of protein/DNA complex at the level of bp, using the fundamental 
concepts of equilbrium statistical mechanics. We have build up from the crystallographic data both for complexed 
and the free state, the free-energy of deformation in six degrees of freedom of a bp in a given DNA 
sequence, treating the bp as a rigid plane. We can thus estimate the change in free energy $\delta 
F_{DNA}^{\alpha}$ of the $\alpha$-th bp upon the complex formation. We further identify from the crystallographic  
data the H-bonded atoms of a given protein residue with the given bp and estimate the energy of binding which is 
essentially the free energy gain $\delta F_{complex}^{\alpha}$ upon complexation, if the complex formation is 
energy dominated, especially at low temperatures. One can thereby estimate the free energy gain of the protein, 
$\delta F_{prot}^{\alpha}$ upon complexation at the $\alpha$-th bp by accounting for the accompanied ion 
and water release. $\delta F_{DNA}^{\alpha}$ sheds some light on the time scale of DNA bp dynamics within the 
elastic approximation. We apply this analysis to the particular complex of TBP/TATA sequence DNA. The TBP/TATA box 
complex is one of the most important and well studied protein/DNA complex 
\cite{jmb1996,Biophysical 1998,jmb2001-16,NAR-2006,ref1-jmb-01}. 
Crystal structures of TBP/TATA box show that TBP binds through minor groove to 
severly deformed\cite{kim} consensus TATA sequence, namely, TATA(T/A)A(T/A)N, where the bases in one of the strands 
have been indicated, (T/A) being either thymine or adenine and N any of the four bases.
The functional groups in the minor groove are incapable to provide enough discriminatory H-bond partners to TBP for specific binding.
An indirect mode of recognition
has been proposed to explain TBP/TATA sequence specific binding. Here the DNA becomes
structurally rigid in severely deformed conformation that allows the TBP to form 
adequate H-bonds. However, it has been found 
from the previous bioinformatics study that DNA becomes more flexible upon protein binding \cite{olson-PNAS}. 
Our analysis on TBP/TATA box complex sheds some light on this specific binding mode which is of profound 
biological interest.

We have taken structures of the available TBP/TATA-sequence complexes solved by x-ray crystallography at a resolution better 
than 2.5$A^{0}$ from the Protein Data Bank (PDB)\cite{PDB} \cite{footnote1}. The temperature of the selected complexes is in 
the range $100-120$K and the pH ranges within $6.5-7.0$. We have taken all such TBP/TATA sequence complexes as 
independent data set. For all the selected complexes TBP binds to a ten bp highly conserved ($>90$\%) DNA CTATAAAAGG sequence in one strand ($5^{\prime}$ to $3^{\prime}$) along with the normal Watson-Crick base pairing 
in the opposite strand. We have taken two consecutive bps in $5^{\prime}$ to $3^{\prime}$ direction  and construct 
a mean axis system with respect to which the bp geometrical parameters are defined\cite{NUPARM}. We define base normal, 
$\vec{N}_{\alpha}$, as the vector normal to mean bp plane defined by all ring atoms of the ${\alpha}$-th base, 
obtained from the coordinate of different atoms listed in the PDB files. We further take the bp normal, 
$\vec{Z}_{\alpha}$, as the mean of $\vec{N}_{\alpha}$ of two paired bases, normalized to unit vector. The bp long axis, 
$\vec{Y}_{\alpha}$, is the vector along the line joining $C1^{\prime}$ atoms of the bps, as shown in
Fig.\ \ref{fig1}(b). The bp short axis, $\vec{X}_{\alpha}$ is the vector normal to both $\vec{Y}_{\alpha}$ and 
$\vec{Z}_{\alpha}$ and pointing towards the major groove side of the bp. The base-base vector, $\vec{M}$, is the 
vector joining the centers of two consecutive bps and mean doublet z-axis, $\vec{Z}_m= \frac{(\vec{X}_1+
\vec{X}_2)\times(\vec{Y}_1+\vec{Y}_2)}{|\vec{X}_1+\vec{X}_2|.|\vec{Y}_1+\vec{Y}_2|}$. The bp parameters have been 
calculated using the relations: Tilt, $\tau=-\sin^{-1} (\vec{Z}_m.\vec{X}_1)$; roll, $\rho=\sin^{-1} 
(\vec{Z}_m.\vec{Y}_1)$; twist, $\omega=\cos^{-1}[(\vec{X}_1\times\vec{Z}_m).(\vec{X}_2\times\vec{Z}_m)]$; 
Shift, $Dx=\vec{M}. \frac{(\vec{X}_1+\vec{X}_2)}{|\vec{X}_1+\vec{X}_2|}$; Slide, $Dy=\vec{M}.\frac{(\vec{Y}_1+\vec{Y}_2)}
{|\vec{Y}_1+\vec{Y}_2|}$ and Rise, $Dz=\vec{M}.\vec{Z}_m$.

\begin{figure}[t]
{\rotatebox{270}{\resizebox{10cm}{8cm}{\includegraphics{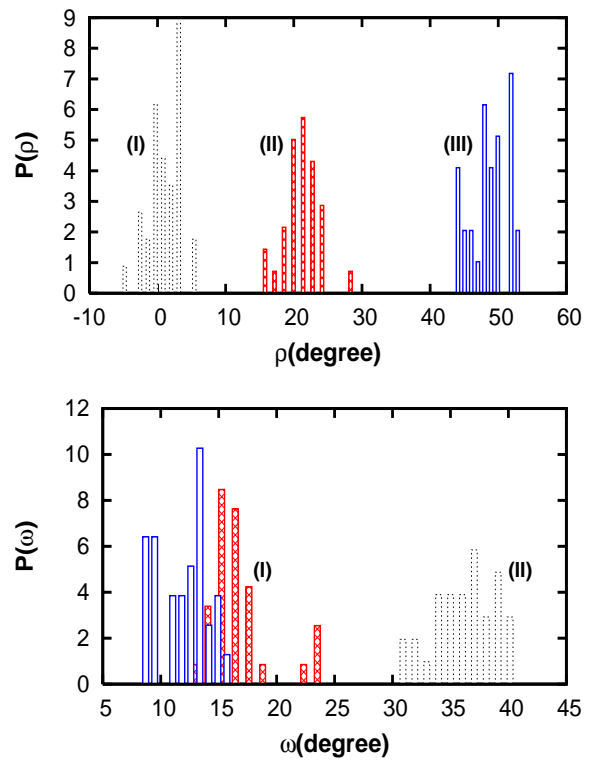}}}}
\caption{(color online) (a) P($\rho$)vs.$\rho$: Set I for the first base pair (fourth and ninth bp being similar).
Set II for sixth bp steps (fifth and seventh being similar) and Set III for second bp steps (third, eighth bp similar trend). 
(b) P($\omega$)vs.$\omega$: Set I for second and sixth base pair step (fifth, seventh bp being similar). Set II for first base pair step (third, fourth, eighth and nineth bp being similar).}
\label{fig2}
\end{figure}

We generate histogram for each of the local parameters of the TATA sequence DNA complexed with TBP. Fig.\ref{fig2}(a) 
shows typical histogram $P(\rho)$ for $\rho$. There are three distinct sets of histogram: Set I shows 
the data for the first bp, set II shows data for the fifth bp and set III shows that for the second bp. 
In set I, the mean value of $\rho$ lies within the range ${0}^\circ-{4}^\circ$ which is similar to that known 
in the free case\cite{free}.  So bps in set I has almost no roll deformation due to complexation. 
On the other hand in set III, where the mean value of $\rho$ is within $48^\circ-52^\circ$, exhibits the 
maximum deformation due to complexation. The mean value of $\rho$ in set II lies within $20^\circ-24^\circ$. 
Fig.\ \ref{fig2}(b) shows the histograms, $P(\omega)$ for $\omega$. Here the mean value of $\omega$ for 
the first bp is $\sim 36^\circ$, comparable to mean $\omega$ in the free case\cite{free}. 
The mean value of $\omega$ corresponding to the second and the sixth bp are $15^\circ$ 
and $18^\circ$ respectively, exhibiting large deformation in $\omega$. The histogram for the 
$\tau$ corresponding to all bp have the mean values similar to the known tilt value in free case 
($0.0^\circ \pm 0.5^\circ$)\cite{free}, having insignificant effect of complexation. 

\begin{figure}[t]
{\rotatebox{270}{\resizebox{10cm}{10cm}{\includegraphics{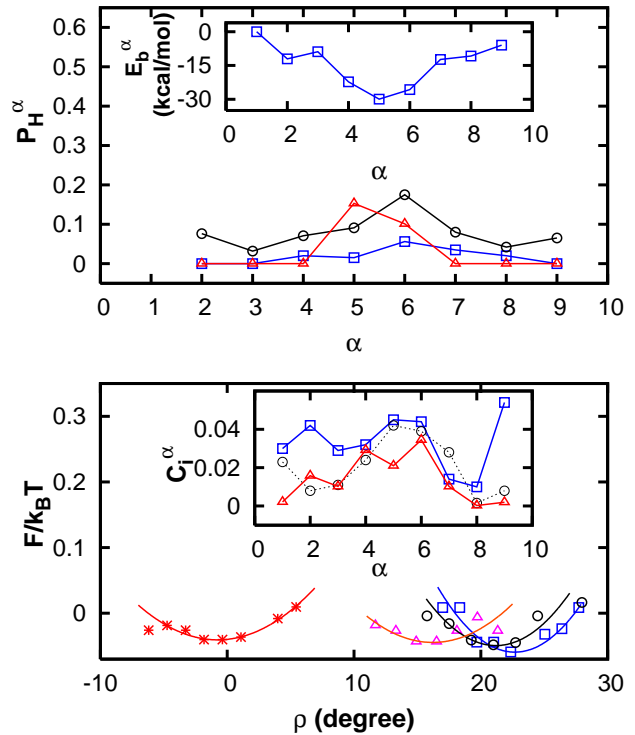}}}}
\caption{(color online) (a) H-Binding probability $P_{H}^{\alpha}$ of sugar ($\square$), 
phosphate($\triangle$) and base ($\circ$) with protein residues plotted against $\alpha$. 
The joining lines are guide to the eyes. 
Inset: Binding energy, $E_{b}^{\alpha}$ vs $\alpha$ plot.
(b) Free energy of deformation corresponding to $\rho$. Leftmost 
data ($*$) for the free case. Rest of the three for bound cases: seventh ($\triangle$) ,sixth ($\square$)
and the fifth ($\circ$) bp respectively. Lines are the best fitted curve. 
Inset: $C_{i}^{\alpha}$ corresponding to the three orientational bp parameters 
plotted against $\alpha$:  $\rho$ ($\square$), $\omega$ ($\triangle$) and
$\tau$ ($\circ$). The joining lines are guide to the eyes.}
\label{fig3}
\end{figure}

We further find the atoms of the protein residues forming H-bonds to the bases in a given bp along with 
the sugar ring and the phosphate groups attached onto it. Here H-bond analysis has been done with 
the help of pyrHBfind software\cite{pyrhb}. 
The binding region is primarily located between the fourth and the seventh bp. 
There are direct 
H-bonding at the fifth and the sixth bp between base and amino acid residues, asparagine and tryptophan. 
Fig.\ \ref{fig3}(a) shows the probability $P_{H}^{\alpha}$ of finding H-bonds through sugar, phosphate and base
in the vicinity of the $\alpha$-th bp.  Maximum number of H-bonding, three with the bases, 
two with the phosphate oxygen and one with the sugar, is observed at the vicinity of fifth bp. There is a substantial number of 
H-bonding through the sixth bp as well. Binding through sugar dominate the between fifth to seventh bp,
the maximum being at the sixth bp. The correlation between the deformation of DNA and the binding pattern is 
better revealed from the inset of Fig.\ \ref{fig3}(a), where the H-bonding energy $E^{\alpha}_{b}$\cite{footnote2} 
between the bound part of the bp and the protein residue, is plotted against bp. We find that the classical 
electrostatic part has dominant contribution to $E^{\alpha}_{b}$ in all the cases. $E^{\alpha}_{b}$ 
has a funnel structure between the third and the seventh bps, having a minimum $\sim -30$ kcal/mole at the fifth bp. 
At the seventh bp, where protein residues do not directly interact with the base but interact with the backbone, 
$E^{\alpha}_{b}$ is relatively small, $\sim -13$ kcal/mole. Similarly, at the second and the nineth bp where the 
phosphate oxygen interacts with polar amino acid residues $E^{\alpha}_{b}$ is $\sim -12$ kcal/mole. Thus the  
base binding contributes an energy $\sim -15$ kcal/mole at the fifth and the sixth bps. This additional energy
can be defined as that due to the specificity of the base binding, the specificity being larger at the fifth bp. 
Note that the binding process makes the bps energetically inhomogeneous which is particularly remarkable for the 
fifth, sixth and the seventh bp, each of them being A. The $E^{\alpha}_{b}$ data corroborates to the 
deformation data in Figs\ \ref{fig2}(a) and (b), namely, the $P(\rho)$ and the $P(\omega)$ data group together for 
the bps in the region of stronger binding. $P(\rho)$ and $P(\omega)$ for the second and the eighth bp are distinct 
from the others, despite having relatively weaker $E^{\alpha}_{b}$. $E^{\alpha}_{b}$ has weak local minima at 
these bps. This indicate that the protein binding may initiate at these two metastable points, having 
strong mechanical deformation. 

We calculate the free energy of deformation per complex at the $\alpha$-th bp, $\beta F^{\alpha}_{i}=
-ln[P^{\alpha}_{i}]$, where $i$ denotes any of the six bp parameter, $\beta=1/k_{B}T$, $k_{B}$ being 
the Boltzmann constant and T the temperature. Fig.\ \ref{fig3}(b) shows $\beta F^{\alpha}(\rho)$ for roll of 
the fifth, sixth and the seventh bp. $\beta F^{\alpha}(\rho)$ has a minimum having a fair degree of
harmonicity, typical for an elastic degree of freedom\cite{chaikin}, with deviations only for $\rho$ 
values far away from the minimum. We find similar trend in all other free energy profiles. The curvature of the 
free energy at the minimum is a measure of the local rigidity corresponding to the bp parameter.
The inset of Fig.\ \ref{fig3}(b) shows the curvature at different bp for the rotational parameters, 
$C_{i}^{\alpha}$, $(i=\rho,\tau,\omega)$. The $C_{i}^{\alpha}$ data show large local 
rigidity for the bp with large $E_{b}^{\alpha}$. One can estimate the frequency of small oscillations about
the equilibirum from the curvature data by having the moment of inertia of a bp with respect to the relevant 
axis of rotation. The time period corresponding to the rotational parmeter ranges between 30-50 ps which is comparable to 
solvation time scale of water molecules in the vicinity of protein/DNA complex\cite{skpal}. 

We also estimate $\delta F_{DNA}^{\alpha}$ and $\delta F_{prot}^{\alpha}$. To this end we selected the fifth and 
the sixth bp, where the base binding gives specific stability to the complex. We compare the results to these of the seventh bp,
having no H-bonding with base and the free $AA$ bp doublet\cite{footnote3}. Fig.\ \ref{fig3}(b) shows the free energy for the 
free case as well.  The second column of Table 1 shows that the change in curvature,
in $\rho$, compared to the free case, is largest at the fifth bp. The difference in free energy 
minima between the complex and the free case gives $\delta F_{DNA,i}^{\alpha}$ for ${\alpha}^{th}$ bp parameter so 
that $\delta F_{DNA}^{\alpha} = \sum_{i} \delta F_{DNA,i}^{\alpha}$. The third column of Table I shows 
$\delta F_{DNA}^{\alpha}$ for $\alpha=5,6$ and $7$. We find that $\delta F_{DNA}^{\alpha}<0$ for all $\alpha$, 
indicating that after complexation the DNA goes to a thermodynamically more favoured state. The fourth column of 
Table I shows the loss of entropy of DNA estimated within the harmonic approximation. The $\alpha$ dependence in 
the entropy loss indicates that different amount of heat generated as the H-bonds are formed at different bp in the
complex. The free energy gain upon complexation at the $\alpha$-th bp is given by: $\delta F_{comp}^{\alpha}
= \delta F_{DNA}^{\alpha}+\delta F_{prot}^{\alpha}+\delta F_{water}^{\alpha}+\delta F_{ion}^{\alpha}$, where
$\delta F_{water}^{\alpha}$ and $\delta F_{ion}^{\alpha}$ are the free energy cost for water and ion released respectively
upon bond formation at the $\alpha$-th bp. We take for low temperature $\delta F_{comp}^{\alpha}\simeq E^{\alpha}_{b}$,
ignoring the entropy effects due to the bond vibrations. However, the water and the ion releases are entropy driven processes.
When TBP binds to the TATA sequence 19 water molecules are released from the interfacial 
sufrace\cite{water-release}. We find that seven phosphate groups in the TATA sequence 
are neutralized by seven positively charged amino acid residues. So we expect that upon complex 
formation on an average 14 ions are released from the interfecial suface. The entropy gain ($\simeq \delta F^{\alpha}_{water}+
\delta F^{\alpha}_{ion}$) due to the displaced ions and water molecules is estimated to be $\sim 38.29 k_{B}T$\cite{footnote4}. 
Thus $\delta F^{\alpha}_{prot}$ can be estimated and shown in the last column of Table I where we observe that the
free energy gain by the protein is maximum at the fifth bp. We thus find that the change in curvature
and the maximum gain in thermodynamic free energy of different components in the TBP/TATA sequence
complex are strongly correlated. The metastable complex through binding at the second and the eighth bp
is stabilized by enhanced base binding at locally more rigid fifth bp. This mechanism holds the key to the indirect 
recognition of the TBP/TATA sequence complex. Earlier attempt \cite{olson-PNAS} might have missed enhanced local
rigidity by taking average over all protein/DNA complexes. 

\begin{table}[h!]
\caption{Thermodynamic data for the selected bp.}
\begin{tabular}{ccccc}
\hline
\hline
${\alpha}$& $\delta F_{DNA}^{\alpha}$& $({C_i}^{\alpha}-{C_i}^{Free})$& $\delta S_{DNA}^{\alpha}$& $\delta F_{prot}^{\alpha}$\\
\hline
\hline
Fifth&-0.23&0.0025&-5.7&-48.9\\
Sixth&-0.25&0.0021&-5.47&-41.26\\
Seventh&-0.23&0.0003&-4.77&-17.12\\
\hline
\end{tabular}
\end{table}

In summary we show here an approach based on the equilibrium statistical mechanics how crystal strucrure data is 
used to obtain the bp-wise stability of a protein/DNA complex. Our analysis shows that the maximum specificity of 
the TBP/TATA sequence complex comes from H-bonding with the fifth bp where the changes in the rigidity
are the maximum, though the binding may proceed from the second and the eigthth bp. Even though our analysis 
has been in the crytal phase, recent studies indicate that the enzymatic properties of protein and nucleic
acids remain intact in the crystal phase\cite{banetal}. Hence, our predictions should hold even for
TBP/TATA complex in the solution phase, pertinent to the in-vivo situations. Our predictions 
may be verified by single molecule experiments where the H-bond forming abilty of different bp 
of the TATA-box sequence can be selectively altered by thio-substitution. We would point out that our analysis is quite general 
and can be applied to any protein/Nucleic acid complex. We shall report such detailed analysis in future publications.

\acknowledgments SS thanks the CSIR for financial support.


\begin{thebibliography}{99}
\bibitem{cell} B. Alberts {\it et al.}, {\it Mol. Biol. of The Cell} (Garland Science, 2002).
\bibitem{caladine} C. R. Calladine {\it et al.}, {\it Understanding DNA} (Elsevier, 2004).
\bibitem{jmb1996} Z. S. Juo {\it et al.}, J. Mol. Biol. {\bf 261}, 239 (1996).
\bibitem{Biophysical 1998} L. Pardo {\it et al.}, Biophys. J. {\bf 75}, 2411, (1998).
\bibitem{jmb2001-16} A. H. Elcock and J. A. McCammon, J. Am. Chem. Soc. {\bf 118}, 3787, (1996).
\bibitem{NAR-2006} H. Faiger {\it et al.}, Nucleic Acids Res. {\bf 34}, 104, (2006).
\bibitem{ref1-jmb-01} G. Guzikevich-Guerstein and Z. Shakked, Nature Struct. Biol. {\bf 3}, 32,(1996).
\bibitem{kim} J. L. Kim {\it et al.}, Nature. {\bf 365}, 520 (1993).
\bibitem{olson-PNAS} W. K. Olson {\it et al.} Proc. Natl. Acad. Sci. USA {\bf 95}, 11163, (1998).
\bibitem{PDB} H. M. Berman {\it et al.} Nucleic Acids Res. {\bf 28}, 235, (2000)
\bibitem{footnote1}
The PDB identifires of the selected complexes are
1qn3, 1qn4, 1qn5, 1qn6, 1qn7, 1qn8, 1qn9, 1qna, 1qnb, 1qnc, 1c9b, 1cdw, 1jfi, 1ngm, 1nvp, 1tgh, 1vol, 1ytb.
\bibitem{NUPARM} D. Bhattachryya and M. Bansal, J. Biomol. Struct. Dynam. {\bf 8}, 539, (1990).
\bibitem{free}
\bibitem{pyrhb} S. Mukherjee {\it et al.} J. Phys. Chem. B {\bf 109}, 10484, (2005).
\bibitem{footnote2}
For ab Initio quantum chemical calculation we use the GAMESS software. The hydrogen atomic position of each of the
case then optimized with HF/6-31G(2p,2d) basis sets, by freezing the heavy atoms. The electrostatic part of the interaction
energy is divided by dielectric constant between protein-DNA interfaces (4.0).
\bibitem{chaikin} P. M. Chaikin and T. C. Lubensky, {\it Principles of Condensed matter Physics} (Cambridge, 1998).
\bibitem{skpal} A. H. Zewail {\it et al.} Proc. Natl. Acad. Sci. USA {\bf 100}, 13746, (2003).
\bibitem{footnote3}
The PDB identifires of the selected complexes are
1en3, 1en8, 1fq2, 1hq7, 1ikk, 1ilc, 1jgr, 1lp7, 1n4e, 1s23, 307d.
\bibitem{water-release} S. Khrapunov and M. Bernowitz, Biophys. J. {\bf 86}, 371, (2004).
\bibitem{banetal} P. Nissen {\it et al.} Science, {\bf 289}, 920 (2000).
\bibitem{footnote4} Entropy of the ion and water estimated via $T\delta S \sim NTR log(\frac{\rho_2}{\rho_1})$, where ${\rho}_2$ is the density of ion/water around DNA \cite{density}, ${\rho}_1$ is the bulk ion/water density and N is the number of ion/water released. 
\bibitem{density} T. V. Chalikian (\it et al.) Biophys. Chem. {\bf 51}, 89, (1994).
\end{thebibliography}
\end{document}